\documentclass[superscriptaddress,aalso receivedpaper,10pt,twocolumn,floats,showpacs,amsmath,amsfonts,amssymb,reprint]{revtex4-2}
\usepackage{graphicx}
\usepackage[colorlinks=true,citecolor=blue]{hyperref}
\usepackage[caption=false]{subfig}
\usepackage{pstricks}
\usepackage{pst-node}
\usepackage{amsbsy}
\usepackage{amsmath}
\usepackage[cal=boondoxo]{mathalfa}
\usepackage{verbatim}
\usepackage{dsfont}

\begin{document}
	
	
	\title{Dissipation enables robust extensive scaling of multipartite correlations}
	\author{Krzysztof Ptaszy\'{n}ski}
   	\email{krzysztof.ptaszynski@ifmpan.poznan.pl}
	\affiliation{Complex Systems and Statistical Mechanics, Department of Physics and Materials Science, University of Luxembourg, 30 Avenue des Hauts-Fourneaux, L-4362 Esch-sur-Alzette, Luxembourg}
	\affiliation{Institute of Molecular Physics, Polish Academy of Sciences, Mariana Smoluchowskiego 17, 60-179 Pozna\'{n}, Poland}
	
	\author{Massimiliano Esposito}
	\email{massimiliano.esposito@uni.lu}
	\affiliation{Complex Systems and Statistical Mechanics, Department of Physics and Materials Science, University of Luxembourg, 30 Avenue des Hauts-Fourneaux, L-4362 Esch-sur-Alzette, Luxembourg}
	
	\date{\today}
	
	\begin{abstract}
We investigate the multipartite mutual information between $N$ discrete-state stochastic units interacting in a network that is invariant under unit permutations. We show that when the system relaxes to fixed point attractors, multipartite correlations in the stationary state either do not scale extensively with $N$, or the extensive scaling is not robust to arbitrarily small perturbations of the system dynamics. In particular, robust extensive scaling cannot occur in thermodynamic equilibrium. In contrast, mutual information scales extensively when the system relaxes to time-dependent attractors (e.g., limit cycles), which can occur only far from equilibrium. This demonstrates the essential role of dissipation in the generation and maintenance of multipartite correlations. We illustrate our theory with the nonequilibrium Potts model.
	\end{abstract}
	
	\maketitle

\textit{Introduction---}Information theory provides a universal framework to characterize bipartite and multipartite correlations in complex physical systems. In the last years, such correlations have received significant attention in the context of quantum many-body systems~\cite{amico2008entanglement,laflorencie2016quantum,de2018genuine}. In particular, many studies were concerned with their universal scaling properties in pure~\cite{abanin2019colloquium}, thermal~\cite{wolf2008area} and nonequilibrium steady states~\cite{gullans2019entanglement,panda2020entanglement,dabruzzo2022logarithmic,caceffo2023entanglement,fraenkel2023extensive,fraenkel2024exact}. In the context of classical stochastic systems, research on information-theoretic correlations has mainly focused on their relation to nonequilibrium thermodynamics (e.g., operation of Maxwell demons)~\cite{sagawa2010generalized,strasberg2013thermodynamics,hartich2014stochastic,horowitz2014thermodynamics,horowitz2015multipartite,parrondo2015thermodynamics,wolpert2019stochastic,ehrich2023energy}. However, most studies focused on relatively small systems. Only recently, information theory began to be used to formalize nonequilibrium stochastic thermodynamics at macroscopic scales~\cite{riccardo2016nonequilibrium,penocchio2022information,freitas2022emergent,FalascoReview}. Among others, this framework was used to characterize the nonequilibrium phase transition associated with the formation of Turing patterns~\cite{falasco2018information}. Recent works have further investigated the system-size scaling of information flow between two subsystems in an autonomous Maxwell demon spanning from microscopic to macroscopic scales. In the electronic realization of this device, made of two coupled CMOS inverters, it was shown that information flow can scale extensively with system size (i.e., the number of electrons) only when dissipation scales supraextensively~\cite{freitas2022maxwell, freitas2023information}. Furthermore, it was shown that this is not possible in a similar setup realized with chemical reaction networks, where the information flow is always subextensive with system size (i.e., the number of molecules)~\cite{bilancioni2023chemical}.

In this Letter, we investigate the scaling properties of multipartite correlations in the stationary state of systems composed of a large number of interacting discrete-state stochastic units. Such systems received a great deal of attention, e.g., in the context of synchronization~\cite{wood2006universality,wood2006critical,wood2007continuous,wood2007effects,fernandez2008athermal,assis2011infinite,assis2012collective,escaff2016synchronization,jorg2017stochastic,herpich2018collective,herpich2019universality,meibohm2024minimum,meibohm2024small,sune2019out,rodrigues2020synchronization,zhang2020energy,guislain2023nonequilibrium,gusilain2024discontinuous,guislain2024collective,guislain2024hidden} or collective enhancement of the performance of heat engines~\cite{golubeva2012efficiency,golubeva2013maximum,vroylandt2018collective,vroylandt2020efficiency,filho2023powerful,mamede2023thermodynamics}. In general, the characterization of information-theoretic quantities in such systems is cumbersome since it requires knowledge of the full probability distribution of the system states whose dimension grows exponentially with the system size. However, the problem greatly simplifies for permutation-invariant systems where all stochastic units are identical and all their pairs are coupled to each other in the same manner~\cite{herpich2018collective,herpich2019universality,meibohm2024minimum,meibohm2024small,golubeva2012efficiency,golubeva2013maximum,vroylandt2018collective,vroylandt2020efficiency,guislain2023nonequilibrium,gusilain2024discontinuous,assis2011infinite,assis2012collective,zhang2020energy,wood2006critical,wood2006universality,wood2007continuous,escaff2016synchronization,jorg2017stochastic,filho2023powerful,mamede2023thermodynamics}. Then, in the large size limit, the system entropy becomes a linear function of the probability distribution of the coarse-grained mesostates of the system~\cite{herpich2020njp,FalascoReview}. As we show later, this enables one to characterize the macroscopic scaling of the multipartite mutual information using the deterministic mean-field equations, which can be easily solved numerically.

Our theory demonstrates that the scaling of multipartite mutual information is always subextensive when the system relaxes to a unique fixed point. Scaling may become extensive when the stationary state of the system corresponds to a probabilistic mixture of several fixed points. However, extensive scaling of correlations is not robust in that case -- it can be suppressed by arbitrarily small perturbations of the system dynamics. A robust extensive scaling of correlations is therefore impossible in thermodynamic equilibrium where only fixed point attractors are allowed~\cite{FalascoReview}. In contrast, a robust extensive scaling of correlations is present when the system relaxes to time-dependent attractors (e.g., limit cycles), which can occur only far from equilibrium~\cite{nicolis1977self}. This shows the essential role of dissipation in maintaining robust multipartite correlations in macroscopic stochastic systems.

\textit{Setup---}We consider systems composed of $N$ units whose dynamics corresponds to classical Markov jumps between $d$ discrete states of each unit. The state of the system is characterized by the microscopic configuration vector $\boldsymbol{\alpha} \equiv \{\alpha_i\}_{i=1}^N$, where $\alpha_i \in \{j \}_{j=0}^{d-1}$ denotes the occupied state of each unit. We further focus on permutation-invariant systems, where units are identical and the stochastic dynamics of each unit is invariant to arbitrary permutations of other units; here we follow a comprehensive treatment of such systems in Ref.~\cite{herpich2020njp}. We also confine our analysis to the stationary state of the system, that is, the asymptotic state obtained in the limit $t \rightarrow \infty$. Due to permutation invariance of the dynamics, this state is also permutation-invariant. As a consequence, it is fully characterized by the probability distribution $\{ p_{\boldsymbol{N}} \}_{\boldsymbol{N}}$, where $p_{\boldsymbol{N}}$ is the sum of probabilities of microscopic configurations $\boldsymbol{\alpha}$ corresponding to the given occupation vector $\boldsymbol{N} \equiv \{ N_j \}_{j=0}^{d-1}$, and $N_j \equiv \sum_{i=1}^N \delta_{\alpha_i,j}$ is the number of units in the state $j$. This distribution is given by the stationary solution $\forall {\boldsymbol{N}}: \: d_t p_{\boldsymbol{N}}=0$ of the mesoscopic master equation
\begin{align} \label{eq:mesmasteq}
d_t p_{\boldsymbol{N}}=\sum_{\lambda} \left[W_\lambda ({\boldsymbol{N}}-\boldsymbol{\Delta}_\lambda) p_{\boldsymbol{N}-{\boldsymbol{\Delta}_\lambda}}- W_\lambda ({\boldsymbol{N}}) p_{\boldsymbol{N}} \right] \,,
\end{align}
where ${\boldsymbol{\Delta}_\lambda}$ is the vector denoting the change of the occupation vector during the Markov jump of type $\lambda$, and $W_\lambda ({\boldsymbol{N}})$ is the transition rate for the jump $\lambda$ with the initial occupation vector $\boldsymbol{N}$. The different jump types $\lambda$ correspond, e.g., to different transitions between states of a single unit. (This kind of description can also be applied to transient dynamics with special initial conditions; see Ref.~\cite{herpich2020njp} for details).

\textit{Multipartite correlations}---The central quantity analyzed in our Letter is the multipartite mutual information~\cite{watanabe1960information}
\begin{align}
I_M \equiv \sum_{i=1}^N S_i - S_\text{tot} \geq 0 \,,
\end{align}
where $S_\text{tot} \equiv-\sum_{\boldsymbol{\alpha}} p_{\boldsymbol{\alpha}} \ln p_{\boldsymbol{\alpha}}$ is the Shannon entropy of the total system, and $S_i \equiv -\sum_{\alpha_i} p_{\alpha_i} \ln p_{\alpha_i}$ is the entropy of the $i$th unit. For permutation-invariant states, the total entropy can be calculated as~\cite{herpich2020njp,FalascoReview}
\begin{align} \label{eq:stot}
    S_\text{tot} = \sum_{\boldsymbol{N}} p_{\boldsymbol{N}} \ln \Omega_{\boldsymbol{N}} -\sum_{\boldsymbol{N}} p_{\boldsymbol{N}} \ln p_{\boldsymbol{N}} \,,
\end{align}
where $\Omega_{\boldsymbol{N}}=N!/(\prod_{j=0}^{d-1} N_j!)$ is the number of microscopic configurations $\boldsymbol{\alpha}$ corresponding to a given occupation vector $\boldsymbol{N}$. We now note that the second term on the r.h.s.\ of Eq.~\eqref{eq:stot} corresponds to the Shannon entropy of the probability distribution $\{p_{\boldsymbol{N}} \}_{\boldsymbol{N}}$. Therefore, it is always nonnegative and scales at most logarithmically with system size: $0 \leq -\sum_{\boldsymbol{N}} p_{\boldsymbol{N}} \ln p_{\boldsymbol{N}} \leq d \ln N$, with the upper bound reached for a uniform probability distribution of mesostates, $p_{\boldsymbol{N}}=1/N^d$. Thus, only the first term can be extensive with system size. For large $N$, we can reexpress this term by replacing the discrete probability distribution $\{p_{\boldsymbol{N}} \}_{\boldsymbol{N}}$ with the continuous probability density $\rho(\boldsymbol{n})=N^{d} p_{\boldsymbol{N}}$, where $\boldsymbol{n} \equiv \boldsymbol{N}/N$ is the normalized occupation vector. We then obtain~\cite{herpich2020njp,FalascoReview}
\begin{align} \label{eq:entrtotclasens}
  \lim_{N \rightarrow \infty} \tfrac{1}{N} S_\text{tot} = \left \langle s(\boldsymbol{n}) \right \rangle = \int \rho(\boldsymbol{n}) s(\boldsymbol{n}) d \boldsymbol{n} \,,
\end{align}
where
\begin{align} \label{eq:intentr}
s(\boldsymbol{n}) \equiv \lim_{N \rightarrow \infty} \tfrac{1}{N} \ln \Omega_{\boldsymbol{N}}=-\sum_{j=0}^{d-1} n_j \ln n_j
\end{align}
is the intensive entropy function, that has the form of a Shannon entropy, and $\langle \cdot \rangle$ denotes the ensemble average. To calculate $S_i$ we note that, due to permutation-invariance, the marginal probability distribution of states $\alpha_i$ is the same for each unit and equal to $\{p_{\alpha_i} \}_{\alpha_i}=\langle \boldsymbol{n} \rangle = \sum_{\boldsymbol{N}} p_{\boldsymbol{N}} \boldsymbol{N}/N$. In the macroscopic limit, we further have $\langle \boldsymbol{n} \rangle = \int \rho(\boldsymbol{n}) \boldsymbol{n} d\boldsymbol{n}$. Therefore, using Eq.~\eqref{eq:intentr}, we have
\begin{align} \label{eq:entrclsingle}
S_i\equiv -\sum_{\alpha_i} p_{\alpha_i} \ln p_{\alpha_i} =s (\langle \boldsymbol{n} \rangle) \,.
\end{align}
Consequently, the intensive (divided by $N$) multipartite mutual information scales asymptotically as \begin{align} \label{eq:asyminf}
\lim_{N \rightarrow \infty} \frac{I_M}{N} = s(\langle \boldsymbol{n} \rangle)-\langle s(\boldsymbol{n} ) \rangle \,.
\end{align}

\textit{Macroscopic limit}---We now derive the main result of our Letter, the explicit expressions for the asymptotic scaling of $I_M/N$ [Eqs.~\eqref{eq:asympclinfdet}--\eqref{eq:asympclinfdet-nonerg}].

To that end, let us first discuss what happens when the thermodynamic limit $N \rightarrow \infty$ is taken before the long time limit. If the probability density $\rho(\boldsymbol{n})$ is initially narrowly focused around some point $\boldsymbol{n}_0$, it will stay narrowly focused around the time evolved state $\boldsymbol{n}_t$ whose evolution is given by deterministic mean-field equation~\cite{FalascoReview,hanggi1982stochastic}
\begin{align} \label{eq:meanfield}
d_t \boldsymbol{n}_t = \sum_{\lambda} \boldsymbol{\Delta}_\lambda w_\lambda (\boldsymbol{n}_t) \,,
\end{align}
where
\begin{align} \label{eq:inttransrate}
w_\lambda(\boldsymbol{n})= \lim_{N \rightarrow \infty} W_\lambda(\boldsymbol{N})/N
\end{align}
is the intensive transition rate. The long-time solutions of Eq.~\eqref{eq:meanfield}, called attractors, correspond to macroscopic phases of the system. The attractors may be either time-independent, called fixed points, or time-dependent, e.g., periodic (called limit cycles) or chaotic. When Eq.~\eqref{eq:meanfield} displays multiple attractors, labeled by $\gamma$, the initial condition $\boldsymbol{n}_0$ will determine which one is eventually reached depending on which basin of attraction it belongs.

We now turn the case which interests us in this Letter, where the long time limit is taken before the large $N$ limit. 
We first consider the situation where the mean-field dynamics has a unique attractor. Then, even when this attractor is time-dependent, the system relaxes to the unique stationary state of the master equation~\eqref{eq:mesmasteq}~\cite{gang1987stationary}. As shown for both regular~\cite{dykman1993stationary,vance1996fluctuations,ge2012landscapes} and chaotic~\cite{nicolis1992comments, xu1993internal,geysermans1993thermodynamic,geysermans1996particle,gaspard2020stochastic} attractors, the probability density $\rho(\boldsymbol{n})$ of the stationary state converges with $N$ to the invariant probability density of the deterministic dynamics, $\lim_{\tau \rightarrow \infty} \tau^{-1} \int_{0}^\tau \delta(\boldsymbol{n}-\boldsymbol{n}_t) dt$, provided that the latter is unique (although in the chaotic case, finite-size effects may be important even for relatively large $N$~\cite{fox1990effect, fox1991amplification, keizer1992reply, wang1997intrinsic, wang1998master}). Consequently, as discussed in Ref.~\cite{gaspard2020stochastic}, the ensemble average of any intensive observable (in particular, any function of $\boldsymbol{n}$) converges to the infinite-time average for the deterministic dynamics. Applying this to Eq.~\eqref{eq:asyminf}, $I_M/N$ can be calculated as
\begin{align} \label{eq:asympclinfdet}
\lim_{N \rightarrow \infty} \frac{I_M}{N} = s(\overline{\boldsymbol{n}_t})-\overline{s(\boldsymbol{n}_t )} \,,
\end{align}
where $\overline{(\cdot )}=\lim_{\tau \rightarrow \infty} \tau^{-1} \int_0^\tau (\cdot ) dt$ denotes the infinite-time average for the deterministic dynamics.

We now generalize this result to the \textit{multistable} case, where mean-field equation~\eqref{eq:meanfield} displays multiple attractors. Then, on short timescales, the system initialized in the basin of attraction of the attractor $\gamma$ will relax to a conditional probability density that will be close to the invariant probability density $\lim_{\tau \rightarrow \infty} \tau^{-1} \int_{0}^\tau \delta(\boldsymbol{n}-\boldsymbol{n}_t^\gamma) dt$, where
$\boldsymbol{n}_t^\gamma$ is the deterministically time-evolved state  confined to that basin. On long timescales, due to rare stochastic jumps between the attractors, each attractor will be occupied with the probability $\mathcal{p}_\gamma$ determined by stationary probability of the master equation~\cite{FalascoReview,ge2012landscapes,kurchan2009equilibrium, hanggi1982stochastic}. Consequently, Eq.~\eqref{eq:asympclinfdet} must be generalized by averaging $\boldsymbol{n}$ and $s(\boldsymbol{n})$ over attractors as well as time,
\begin{align} \label{eq:asympclinfdet-nonerg}
\lim_{N \rightarrow \infty} \frac{I_M}{N} = s\left(\sum_\gamma \mathcal{p}_\gamma \overline{\boldsymbol{n}^\gamma_t} \right)-\sum_\gamma \mathcal{p}_\gamma \overline{s \left (\boldsymbol{n}_t^\gamma \right)} \,.
\end{align}
We note that $\mathcal{p}_\gamma$ can be determined by solving the stationary distribution of a coarse-grained master equation describing transitions between attractors using the escape rates from each attractor (see Eq.~(217) in Ref.~\cite{FalascoReview}); for methods to determine these rates, see Refs.~\cite{hanggi1984bistable,dykman1994large, zakine2023minimum}. 

\textit{Conditions of robust extensive scaling of $I_M$}---Let us now discuss the implications of Eqs.~\eqref{eq:asympclinfdet}--\eqref{eq:asympclinfdet-nonerg} for the scaling of mutual information $I_M$. We observe that $I_M$ does not scale extensively with system size (i.e., $\lim_{N\to \infty} I_M/N=0$) when the system relaxes to a single fixed point, because in that case $s(\overline{\boldsymbol{n}_t})=\overline{s(\boldsymbol{n}_t )}$. Otherwise, the mutual information scales extensively and is strictly positive ($\lim_{N\to \infty} I_M/N > 0$). Positivity is ensured by Jensen's inequality $s(\langle \boldsymbol{n} \rangle) \geq \langle s(\boldsymbol{n}) \rangle $ that holds since $s(\boldsymbol{n})$ is a concave function. Consequently, the extensive scaling of $I_M$ is possible in two scenarios: (I) the system relaxes to a time-dependent attractor or (II) several fixed points coexist, i.e.,  several probabilities $\mathcal{p}_\gamma$ are of order 1.

However, scenario (II) (coexistence of fixed points) can occur only for some well-tuned models, because (for $N \rightarrow \infty$) usually only a single attractor (the ``most likely'' one) is occupied with probability $\mathcal{p}_\gamma = 1$ and determines the macroscopic state of the system, while the probabilities of other attractors are exponentially suppressed with $N$~\cite{hanggi1982stochastic,kurchan2009equilibrium,guislain2023nonequilibrium,gusilain2024discontinuous,ge2009thermodynamic,vellela2009stochastic,zakine2023minimum,dykman1994large,FalascoReview}. As a result, this scenario is not robust to small perturbations of the system dynamics. By this we mean that the coexistence of fixed points becomes suppressed (i.e., one of the attractors becomes occupied with probability $\mathcal{p}_\gamma=1$) under certain perturbations of transition rates  $W_{\lambda}(\boldsymbol{N}) \rightarrow W_{\lambda}(\boldsymbol{N})+\varepsilon G_{\lambda}(\boldsymbol{N})$, with arbitrarily small $\varepsilon$ and finite $G_{\lambda}(\boldsymbol{N})$. Physically, such perturbation may be related to the change of some system parameters. For example, in the ferromagnetic phase of the Ising model, two fixed points with opposite magnetization are occupied with probability $\mathcal{p}_\gamma=1/2$ at zero magnetic field, while any finite magnetic field makes one of them occupied with probability $\mathcal{p}_\gamma=1$. More generally, the coexistence of fixed points is observed at discontinuous phase transition points separating two macroscopic phases, but is suppressed by any perturbation of the system parameters that moves it away from the phase transition point~\cite{hanggi1982stochastic,guislain2023nonequilibrium,gusilain2024discontinuous,ge2009thermodynamic,vellela2009stochastic,zakine2023minimum,dykman1994large}.

In contrast, the extensive scaling in scenario (I) can be robust, since -- as well known in the dynamical systems theory -- there are many examples of time-dependent attractors which are \textit{structurally stable}, i.e., whose existence is robust to small perturbations of the transition rates (e.g., hyperbolic limit cycles~\cite{thompson2002nonlinear,anishchenko2014deterministic} or certain chaotic attractors~\cite{guckenheimer1979structural,tucker1999lorenz,gonchenko2021wild,karateskaia2025robust}). Crucially, such attractors can occur only out of equilibrium, which implies that robust extensive scaling of correlations can only occur in the presence of energy dissipation. Conversely, robust extensive scaling of multipartite correlations cannot occur in thermal equilibrium, where only fixed point attractors are allowed (as proven in Ref.~\cite{FalascoReview}).

\begin{figure}
    \centering
    \includegraphics[width=0.9\linewidth]{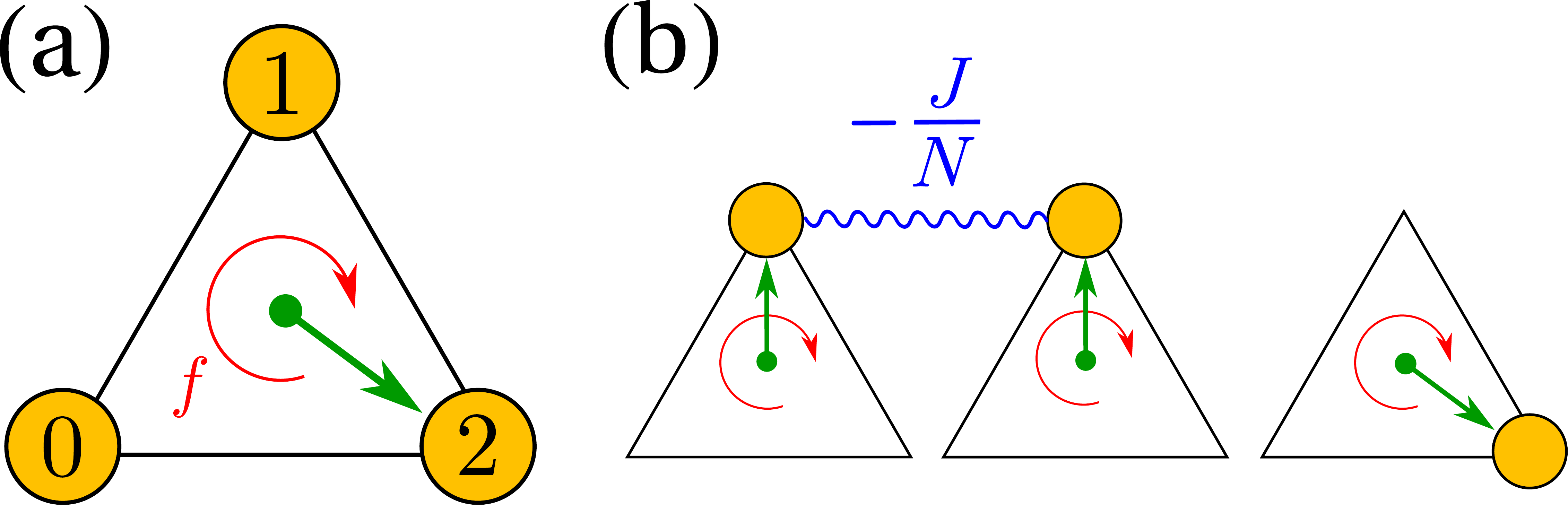}
    \caption{(a) Single unit with 3 states (orange bullets), with a green arrow pointing to the occupied state. The nonconservative force $f$ drives transitions in the clockwise direction. (b) The system energy is reduced by $J/N$ for each pair of units occupying the same state.
    }
    \label{fig:schem}
\end{figure}
\textit{Example: nonequilibrium Potts model---}We illustrate our theory with the nonequilibrium Potts model (Fig.~\ref{fig:schem}), a minimum thermodynamically consistent model of synchronization~\cite{herpich2018collective,herpich2019universality,meibohm2024minimum,meibohm2024small}. It consists of $N$ $d$-state units coupled via all-to-all ferromagnetic interaction so that the system energy is reduced by $J/N$ for every pair of units occupying the same state. Consequently, the system energy $E_{\boldsymbol{N}}$ is solely a function of the occupation vector $\boldsymbol{N}$,
\begin{align}
E_{\boldsymbol{N}}=\sum_{j=0}^{d-1} \epsilon_j N_j-\frac{J}{2 N} \left(\boldsymbol{N} \cdot \boldsymbol{N}-N \right) \,,
\end{align}
where $\epsilon_j$ is the energy of the unit state $j$. It is assumed that only transitions between adjacent states of the units (i.e., $j$ and $j \pm 1$, with $j$ defined modulo $d$) are allowed. The system is also driven by a nonconservative force such that the transition $j \rightarrow j \pm 1$ is associated with the flow of energy $\pm f$ to the environment. 

The dynamics of the model is described by the master equation~\eqref{eq:mesmasteq} with the jump of type $\lambda=(j,j \pm 1)$ corresponding to the transition $j \rightarrow j \pm 1$, and $\boldsymbol{\Delta}_{(j, j \pm 1)}= \{ \delta_{j \pm 1,k}-\delta_{jk} \}_{k=0}^{d-1}$. To provide consistency with the laws of thermodynamics, the transition rates must obey the local detailed balance condition~\cite{herpich2020njp,FalascoReview}
\begin{align}
\ln \frac{W_{(j ,j\pm 1)} (\boldsymbol{N})}{W_{(j \pm 1, j)} (\boldsymbol{N'})}=\beta \left( \pm f +F_{\boldsymbol{N}}-F_{\boldsymbol{N'}} \right) \,,
\end{align}
where $\beta=1/(k_B T)$ is the inverse temperature of the environment$, \boldsymbol{N'}=\boldsymbol{N}+\boldsymbol{\Delta}_{(j,j \pm 1)}$, and $F_{\boldsymbol{N}}=E_{\boldsymbol{N}}-\beta^{-1} \ln \Omega_{\boldsymbol{N}}$ is the free energy of the mesostate $\boldsymbol{N}$. In absence of nonconservative force, the system relaxes to the equilibrium state $p_{\boldsymbol{N}}^\text{eq} \propto e^{-\beta F_{\boldsymbol{N}}}$. 
Specifically, following Refs.~\cite{herpich2018collective,herpich2019universality,meibohm2024minimum,meibohm2024small}, we define the transition rates, and the intensive rates~\eqref{eq:inttransrate}, using the Arrhenius rate model
\begin{subequations}
\begin{align} 
W_{(j, j \pm 1)} (\boldsymbol{N})&= \Gamma N_j e^{\beta[\pm f-\epsilon_{j \pm 1}+\epsilon_j +J(N_{j \pm 1}-N_j+1)/N]/2} \,, \\
w_{(j,j \pm 1)} (\boldsymbol{n})&= \Gamma n_j e^{\beta[\pm f-\epsilon_{j \pm 1}+\epsilon_j+J(n_{j \pm 1}-n_j)]/2} \,,
\end{align}
\end{subequations}
where $\Gamma$ is the kinetic rate.

\begin{figure}
    \centering
    \includegraphics[width=0.9\linewidth]{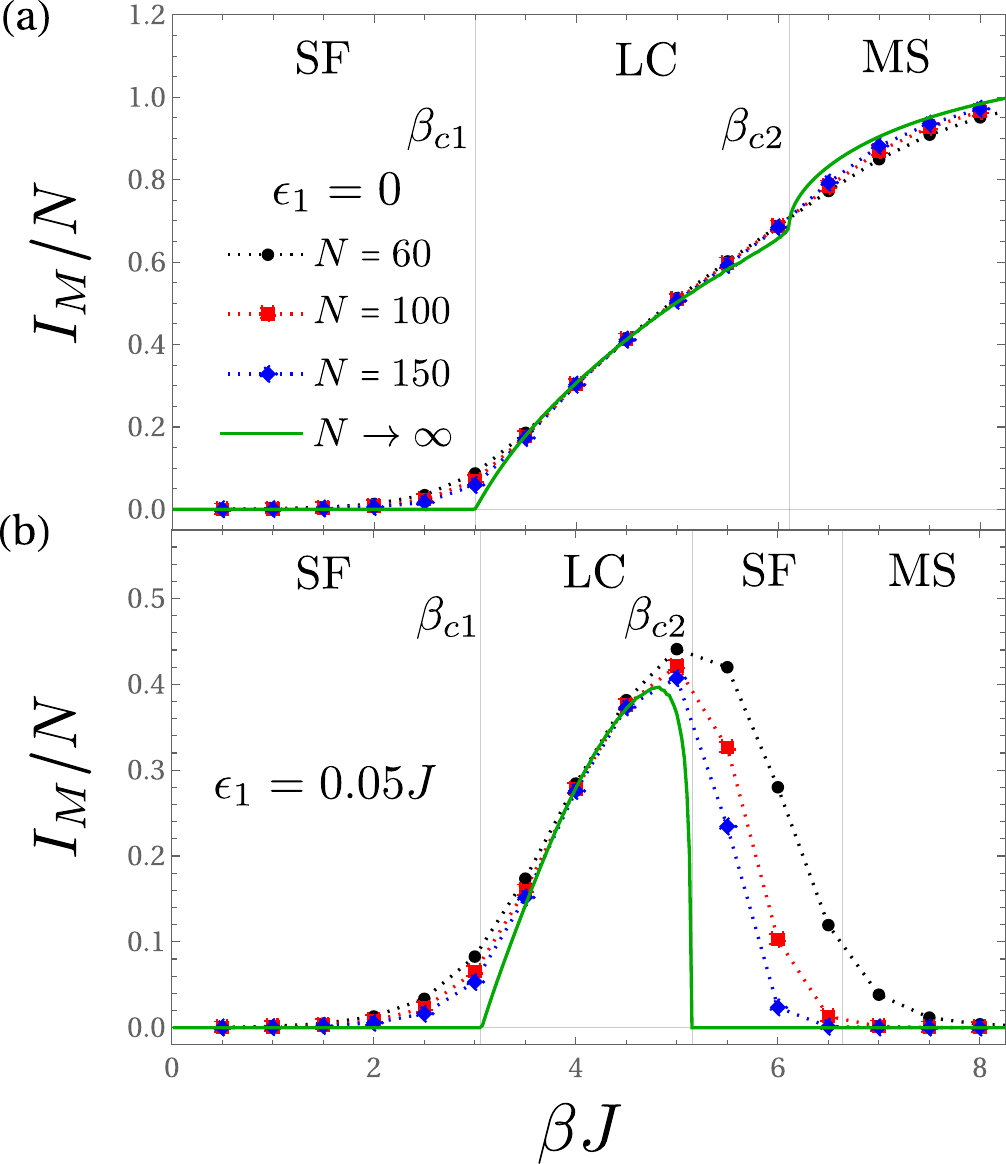}
    \caption{The intensive multipartite mutual information $I_M/N$ for $\epsilon_0=\epsilon_2=0$ and $\epsilon_1=0$ (a), $\epsilon_1=0.05 J$ (b). The solid green line represents the predictions of our theory and dots represent the master equation results for different system sizes $N$. Dotted lines are added for eye guidance. The acronyms in the top denote the single fixed point (SF), limit cycle (LC), and multistable (MS) phases. Parameter: $f=J$.}
    \label{fig:mutinf}
\end{figure}
We now focus on the case $d=3$. First, as in Refs.~\cite{herpich2018collective,herpich2019universality,meibohm2024minimum,meibohm2024small}, we consider a system with cyclic symmetry (i.e., symmetry with respect to the cyclic permutation of $\boldsymbol{N}$) by taking $\epsilon_0=\epsilon_1=\epsilon_2=0$. In this case, when increasing $\beta$, the system exhibits a sequence of two continuous nonequilibrium phase transitions between three distinct phases; see Ref.~\cite{herpich2018collective} for a detailed analysis. Below $\beta_{c1}=3J^{-1}$, the system has a single fixed point (SF phase). Between $\beta_{c1}$ and $\beta_{c2} \approx 6.11 J^{-1}$ the system exhibits a unique limit cycle attractor (LC phase). Finally, at $\beta_{c2}$, the system enters the multistable phase (MS) with three stable fixed points via the infinite period (IP) bifurcation. Due to the cyclic symmetry of the system, these fixed points are occupied with equal probabilities $\mathcal{p}_\gamma=1/3$. Consequently, in the SF and LC phases we calculate $I_M/N$ using Eq.~\eqref{eq:asympclinfdet}, while for the MS phase we use Eq.~\eqref{eq:asympclinfdet-nonerg}. As shown in Fig.~\ref{fig:mutinf}~(a), the intensive mutual information $I_M/N$ appears at the Hopf bifurcation and then grows monotonically with $\beta$, exhibiting a nonanalytic behavior at the IP bifurcation. This behavior is confirmed by the master equation results for finite system sizes, which quantitatively agree with the predictions of our theory, especially in the limit cycle phase.

\begin{figure}
    \centering
\includegraphics[width=0.95\linewidth]{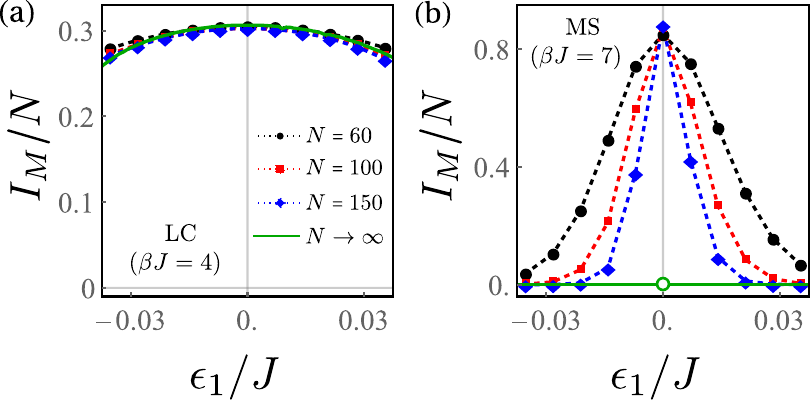}
    \caption{$I_M/N$ as a function of $\epsilon_1=-\epsilon_2$ in (a) limit cycle (LC) phase with $\beta J=4$ and (b) multistable (MS) phase with $\beta J=7$. Parameters: $f=J$, $\epsilon_0=0$. Symbols as in Fig.~\ref{fig:mutinf}; note that the results denoted by the green line in (b) apply to all $\epsilon_1$ apart from $\epsilon_1=0$.
    }
    \label{fig:mutinf-epsilon}
    \end{figure}

To verify the robustness of extensive scaling of $I_M$, inspired by Ref.~\cite{assis2012collective}, we now break the cyclic symmetry by taking $\epsilon_1 >0$. Then, the transition $0 \rightarrow 1$ is suppressed, and the units tend to predominantly occupy the state 0. As before, the system enters the limit cycle phase at $\beta_{c1} \approx 3.05 J^{-1}$ due to the Hopf bifurcation. However, the IP bifurcation at $\beta_{c2} \approx 5.15J^{-1}$ brings the system again to the SF phase, rather than to the multistable phase. Such a reentry into the same phase is called the reentrant phase transition~\cite{castro1995reentrance,radicchi2006reentrant}. As a result, as illustrated in Fig.~\ref{fig:mutinf}~(b), $I_M/N$ exhibits a  nonmonotonic behavior: it appears at the Hopf bifurcation and vanishes at the IP bifurcation. When $\beta$ increases further, the system becomes multistable, with a second (third) stable fixed point generated by a saddle-node bifurcation~\cite{assis2012collective} at $\beta \approx 6.64 J^{-1}$ ($\beta \approx 6.66 J^{-1}$). However, $I_M/N$ still vanishes for $N \rightarrow \infty$, because a single attractor, corresponding to a predominant occupation of state 0, tends to be occupied with probability $\mathcal{p}_\gamma =1$. The nonmonotonic behavior of $I_M/N$ is reproduced by the master equation results, which gradually approach the predictions of our theory for increasing $N$. In the limit cycle phase, the results agree quantitatively already for a relatively small $N=60$. This shows that the extensive scaling of multipartite mutual information is robust to perturbations breaking the cyclic symmetry for limit cycle attractors, while it is not robust for fixed point attractors; the former results from hyperbolicity of the limit cycle, which makes its existence robust to perturbations (see Appendix~\ref{app:hyperb}).

We illustrate those observations further by considering the dependence of $I_M/N$ on the magnitude of perturbation from the cyclic symmetry. We parameterize this magnitude by $\epsilon_1$, taking $\epsilon_0=0$ and $\epsilon_2=\mathcal{a} \epsilon_1$. In Fig.~\ref{fig:mutinf-epsilon} we present the results for $\mathcal{a}=-1$. As shown, in the LC phase, for both $\epsilon_1=0$ and $\epsilon_1 \neq 0$, $I_M/N$ converges with $N$ to a finite value that agrees with the predictions of our theory. In contrast, in the MS phase, $I_M/N$ is suppressed with $N$ for $\epsilon_1 \neq 0$, so that it becomes increasingly narrowly peaked around
the cyclic symmetry point. As discussed in the Appendix~\ref{app:perturb}, the same behavior is observed also for other parameters $\mathcal{a}$, except for some peculiar values, for which the energy perturbation goes along the discontinuous phase transition line in $(\epsilon_1,\epsilon_2 )$ plane. This confirms that the extensive scaling of $I_M$ is robust to perturbations in the LC phase, but not in the MS phase.

\textit{Final remarks---}We note that a robust extensive scaling of multipartite correlations in time-dependent attractors is accompanied by an extensive scaling of the energy dissipation~\cite{FalascoReview}. This contrasts with bipartite correlations between two macroscopic systems, whose extensive scaling was shown to require dissipation that scales superextensively with system size~\cite{freitas2022maxwell,freitas2023information}. An open question is whether our conclusions concerning the conditions for an extensive scaling of correlations hold beyond the permutation-invariant systems that we considered in our study.
For finite-dimensional lattice models~\cite{wood2006universality, wood2006critical, assis2011infinite, sune2019out, guislain2024collective,rodrigues2020synchronization,fernandez2008athermal,filho2023powerful} an interesting direction may be the information-theoretic characterization of complex spatiotemporal patterns (such as waves), which do not exist in permutation-invariant systems~\cite{escaff2014arrays,noguchi2024cycling,noguchi2024spatiotemporal,noguchi2024spatiotemporalpatternsactivefourstate}. Disordered systems~\cite{wood2007effects,guislain2024hidden} should also be investigated. Since these systems can give rise to complex free energy landscape at equilibrium with an extensive number of degenerate minima (e.g. spin glasses~\cite{castellani2005spin}), different scaling of suitably defined mutual information may be expected.

Beyond demonstrating a fundamental link between dissipation and correlations, our work has practical relevance. Among others, in the vein of Ref.~\cite{ameri2015mutual}, the multipartite mutual information $I_M$ can be used as a universal (model-independent) order parameter of synchronization in all-to-all coupled networks consisting of a large number of oscillators~\cite{herpich2018collective,herpich2019universality,meibohm2024minimum,meibohm2024small,guislain2023nonequilibrium,gusilain2024discontinuous,assis2011infinite,assis2012collective,zhang2020energy,wood2006critical,wood2006universality,wood2007continuous,escaff2016synchronization,jorg2017stochastic}. Equation~\eqref{eq:asympclinfdet} enables a practical calculation of this quantity using mean-field dynamics. Additionally, as demonstrated in the Appendix~\ref{app:perturb}, $I_M$ can also be used as a witness of emergent discontinuous phase transitions in finite-size systems.

\begin{acknowledgments}
The authors thank Vasco Cavina for inspiring discussions. K.P. acknowledges the financial support of the National Science Centre, Poland, under the project No.\ 2023/51/D/ST3/01203.

\textit{Data availability}---Wolfram Mathematica notebooks used to obtain the numerical results and the data used in the figures are available at the following DOI: 10.5281/zenodo.13939399.
\end{acknowledgments}

\appendix

\begin{center}
  \large \bf End Matter
\end{center}

\section{Structural stability of limit cycles}\label{app:hyperb}

In the main text, we noted that the existence of hyperbolic limit cycles is robust to small perturbations of the transition rates $W_\lambda(\boldsymbol{N})$ [and thus, the intensive transition rates $w_\lambda(\boldsymbol{\boldsymbol{n}})$ appearing in Eq.~\eqref{eq:meanfield}]. In dynamical systems theory, this feature is called \textit{structural stability}~\cite{thompson2002nonlinear,anishchenko2014deterministic}. We also stated that the limit cycle in the Potts model is hyperbolic, and thus robust. Here we briefly define how the hyperbolicity of the limit cycle is defined. To that end, we first define the Jacobian of the deterministic dynamics evaluated around the time evolved state $\boldsymbol{n}_t$,
\begin{align}
\mathbb{J}(\boldsymbol{n}_t)=\left[ \partial_{n_k} \dot{n}_j (\boldsymbol{n}_t) \right ]_{1 \leq j,k \leq d-1} \,,
\end{align}
where $\dot{n}_j(\boldsymbol{n})=\sum_{\lambda} \Delta_{\lambda,j} w_\lambda(\boldsymbol{n})$ is the deterministic rate of change of the element $n_j$ given by Eq.~\eqref{eq:meanfield}. We note that the Jacobian has dimension $d-1$ rather than $d$, because, due to the conservation law $\sum_{j=0}^{d-1} n_j=1$, the system dynamics is characterized by $d-1$ independent variables $ \{n_1,\ldots,n_{d-1} \}$, with $n_0=1-\sum_{j=1}^{d-1} n_j$. We then focus on a situation where the system exhibits a time-periodic limit cycle solution with $\boldsymbol{n}_{t}=\boldsymbol{n}_{t+T}$, where $T$ is the oscillation period. We consider the time-evolution of the fundamental matrix $\mathbb{M}(t)$,
\begin{align}
d_t \mathbb{M}(t)=\mathbb{J}(\boldsymbol{n}_t) \mathbb{M}(t) \,,
\end{align}
with $\mathbb{M}(0)=\mathds{1}$. The hyperbolicity of the limit cycle is then determined by the eigenvalues of the monodromy matrix $\mathbb{M}(T)$, called the \textit{Floquet multipliers}: the cycle is hyperbolic if a single Floquet multiplier equals $1$, while all others have moduli different from 1. In particular, for attractive limit cycles, the latter multipliers have moduli smaller than 1. We verified numerically that this is true for the Potts model considered in the main text, which confirms that the existence of the limit cycle is robust to small perturbations of the dynamics.

\begin{figure}
    \centering
\includegraphics[width=0.95\linewidth]{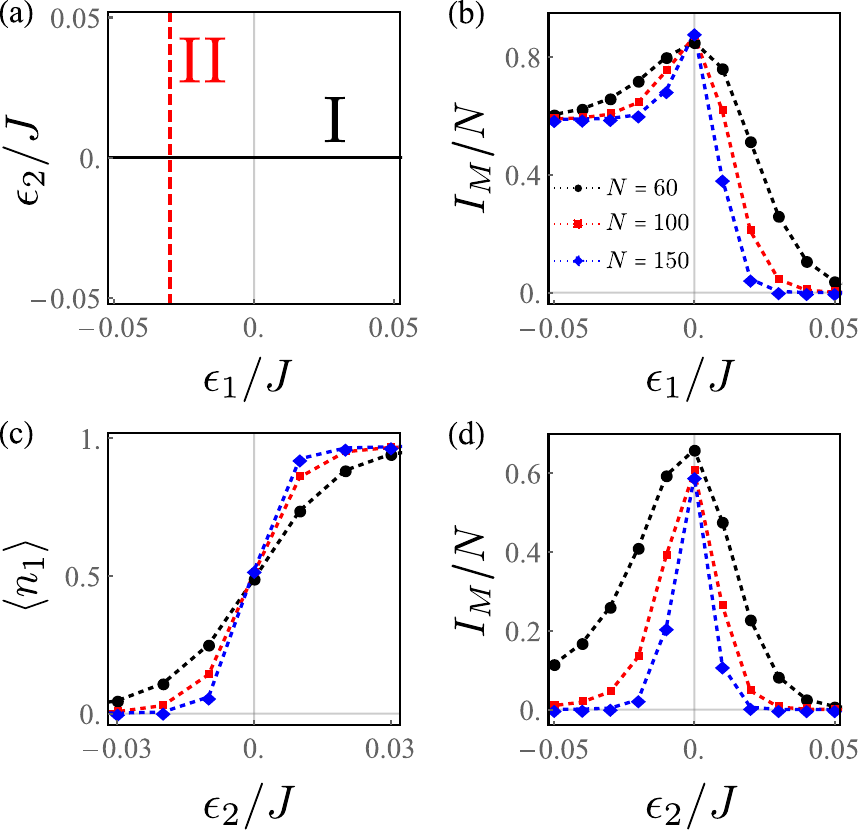}
    \caption{(a) Lines in $(\epsilon_1,\epsilon_2)$ plane representing the considered energy perturbations. (b) $I_M/N$ as a function of $\epsilon_1$ for $\epsilon_2=0$ [along line I in (a)]. (c,d) $\langle n_1 \rangle$ and $I_M/N$ as a function of $\epsilon_2$ for $\epsilon_1=-0.03J$ [along line II in (a)]. Parameters: $f=J$, $\epsilon_0=0$.}
    \label{fig:appendix}
    \end{figure}

\section{Energy perturbations in the multistable phase} \label{app:perturb}

In this Appendix, we analyze in more detail the influence of perturbations of energies $\epsilon_j$ on the scaling of $I_M$ in the multistable phase of the Potts model. Specifically, we consider the dependence of $I_M/N$ on $\epsilon_1,\epsilon_2$, setting $\epsilon_0=0$ as a reference. We recall that in Fig.~\ref{fig:mutinf-epsilon} we considered a specific parameterization $\epsilon_2=-\epsilon_1$. For such a case, we observed that $I_M/N$ is suppressed with $N$ for $\epsilon_1 \neq 0$. In fact, we observed such a behavior for most parametrizations $\epsilon_2=\mathcal{a} \epsilon_1$ (not shown). However, there are some peculiar exceptions for $\mathcal{a} \approx \{0,1,\pm \infty \}$.

As an example of such exception, let us consider the case of $\mathcal{a}=0$ ($\epsilon_2=0$). This corresponds to the sweep of $\epsilon_1$ along line I in Fig.~\ref{fig:appendix}~(a). The behavior of $I_M/N$ is presented in Fig~\ref{fig:appendix}~(b). As in the case analyzed in Fig.~\ref{fig:mutinf-epsilon}, $I_M/N$ is suppressed with $N$ for $\epsilon_1>0$. However, in contrast to the previously described behavior, $I_M/N$ does not vanish for $\epsilon_1<0$.

To explain this behavior, we recall that $I_M$ can scale extensively when several fixed points are occupied with finite probabilities $\mathcal{p}_\gamma$. Such a situation occurs, for example, at discontinuous phase transition points separating macroscopic phases corresponding to different fixed points~\cite{hanggi1982stochastic,guislain2023nonequilibrium,gusilain2024discontinuous,ge2009thermodynamic,vellela2009stochastic,zakine2023minimum,dykman1994large}. In the setup considered, in the multistable phase, the system exhibits three stable fixed points. Consequently, the phase diagram in the $(\epsilon_1,\epsilon_2)$ plane consists of three macroscopic phases corresponding to these fixed points. These regions are separated by three discontinuous phase transitions lines in the $(\epsilon_1,\epsilon_2)$ plane. Due to symmetry reasons, these line emanate from the cyclic symmetry point $\epsilon_1=\epsilon_2=0$, where all three fixed points coexist with the same probability. Thus, the apparent non-vanishing of $I_M/N$ for $\epsilon_2=0$ and $\epsilon_1<0$ suggests that, for those parameter values, the system is at (or at least close to) one of discontinuous phase transition lines in $(\epsilon_1,\epsilon_2$) plane.

To confirm that, in Fig.~\ref{fig:appendix}~(c,d) we analyze the effect of perturbation of  $\epsilon_2$ for a fixed $\epsilon_1=-0.03J$. This corresponds to the sweep of $\epsilon_2$ along line II in Fig.~\ref{fig:appendix}~(a), which can be expected to be perpendicular to the discontinuous phase transition line. In fact, as shown in Fig.~\ref{fig:appendix}~(c), the normalized occupancy of the state $j=1$, labeled $\langle n_1 \rangle$, exhibits a rapid jump at $\epsilon_2 \approx 0$, which becomes increasingly sharp with increasing $N$. Analogously, one observes also a rapid decrease of $\langle n_2 \rangle$ (not shown).  This witnesses the presence of discontinuous phase transition between fixed points with dominant occupations of states $j=1$ and $j=2$.  In agreement with our theory, $I_M/N$ is suppressed with $N$ on both sides of the phase transition, so that it becomes increasingly narrowly peaked around the phase transition point [Fig.~\ref{fig:appendix}~(d)]. The magnitude of this peak is around $I_M/N \approx 0.6$. This approximately corresponds to $I_M/N \approx 0.605$ calculated at $\epsilon_2=0$ using Eq.~\eqref{eq:asympclinfdet-nonerg}, assuming equal probabilities $\mathcal{p}_\gamma=1/2$ of the fixed points separated by the phase transition.

We thus conclude that the extensive scaling of $I_M/N$ in the multistable case is suppressed for most points $(\epsilon_1,\epsilon_2)$, apart from those located at discontinuous phase transition lines in the $(\epsilon_1,\epsilon_2)$ plane. This further suggests that the analysis of $I_M$ may serve the detection of emergent discontinuous phase transitions in simulations of finite-size systems.

\bibliography{bibliography}	
	
\end{document}